\documentclass[onecolumn]{mn2e}
\newif\ifAMStwofonts

\def\pmb#1{\mbox{\boldmath$#1$}}
\def\gtsim {>\kern-1.2em\lower1.1ex\hbox{$\sim$}}
\def\ltsim {<\kern-1.2em\lower1.1ex\hbox{$\sim$}}
\def\gtsim {>\kern-1.2em\lower1.1ex\hbox{$\sim$}}
\def\ltsim {<\kern-1.2em\lower1.1ex\hbox{$\sim$}}
\def\ref{\hangindent=1pc \hangafter=1 \noindent}
\usepackage{epsfig}
\begin{document}

\title[$R$ modes of slowly pulsating B stars]{$R$ modes of slowly pulsating B stars}

\author[U. Lee]{Umin Lee$^1$ \thanks{E-mail:
lee@astr.tohoku.ac.jp} \\$^1$Astronomical Institute, Tohoku University,
Sendai, Miyagi 980-8578, Japan}

\date{Typeset \today ; Received / Accepted}
\maketitle

\begin{abstract} Slowly pulsating B (SPB) stars are $g$ mode pulsators 
in main-sequence stages with mass ranging from
$M\sim3M_\odot$ to $\sim 8M_\odot$.
In this paper, we examine pulsational stability of low $m$ $r$ modes
in SPB stars by calculating
fully nonadiabatic oscillations of uniformly rotating stars, where
$m$ is an integer representing the azimuthal wave number around the rotation axis.
$R$ modes are rotationally induced, non-axisymmetric, oscillation modes, whose
oscillation frequency strongly depends on the rotation frequency $\Omega$ of the star.
They are conveniently classified by using two
integer indices $m$ and $l^\prime\ge |m|$ that define the asymptotic oscillation frequency
$2m\Omega/[l^\prime(l^\prime+1)]$ in the limit of $\Omega\rightarrow 0$.
We find low $m$, high radial order, odd $r$ modes with $l^\prime=m$ in SPB stars
are excited by the same iron opacity bump mechanism that excites 
low frequency $g$ modes of the variables,
when the rotation frequency $\Omega$ is sufficiently high.
No even $r$ modes with low $m$ are found to be pulsationally unstable.
Since the surface pattern of the temperature perturbation of odd modes 
is antisymmetric about the equator of the star, 
observed photometric amplitudes caused by the unstable
odd $r$ modes with $l^\prime=m$ are strongly dependent 
on the inclination angle between the axis of rotation
and the line of sight.
Applying the wave-meanflow interaction formalism to nonadiabatic $r$ modes in rapidly
rotating SPB models, we find that because of the $r\phi$ component of the Reynolds stress and
the radial transport of the eddy fluctuation of density in the rotating star,
the surface rotation is accelerated by
the forcing due to the low $l^\prime=m$ unstable $r$ modes.
We suggest that the amount of angular momentum redistribution in the surface region
of the stars can be comparable to that needed to sustain decretion discs found in Be systems.

We make a brief comparison between nonadiabatic $r$ mode calculations done 
with and without the traditional approximation.
In the traditional approximation,
the local horizontal component of the rotation vector $\pmb{\Omega}$
is ignored in the momentum conservation equation, which makes it possible 
to represent the angular dependence of an oscillation mode using a single Hough function.
We find that the oscillation frequencies of low $m$ $r$ modes 
computed with and without the traditional approximation are qualitatively in good agreement.
We also find that the pulsational instability of $r$ modes in the traditional approximation 
appears weaker than that without the approximation.

\end{abstract}

\begin{keywords}
stars: oscillations -- stars : rotation 
\end{keywords} 


\maketitle  

\section{Introduction}

Slowly pulsating B stars (hereafter SPB stars) are g mode pulsators
(e.g., Waelkens 1991, Waelkens et al 1998).
Low frequency $g$ mode oscillations in SPB stars are excited by the iron opacity bump mechanism, 
which also excites high frequency $p$ modes in $\beta$ Cephei stars 
(Dziembowski \& Pamyatnykh 1993; Dziembowski, Moskalik, \& Pamyatnykh 1993; 
Gautschy \& Saio 1993, see also Kiriakidis, ElEid, \& Glatzel 1992).
Savonije (2005) and Townsend (2005b) have recently shown that low $m$ $r$ modes in SPB stars are also
excited by the same opacity bump mechanism that excites $g$ modes, if rapid rotation is assumed.
Although SPB stars are not necessarily rapid rotators, a fraction of SPB stars
are rapidly rotating (e.g., Aerts et al 1999).
Rapidly rotating SPB stars will be good observational targets for both $g$ mode and
$r$ mode oscillations, which can be used as a probe to investigate the interior structure of 
the variables.

$R$ modes are rotationally induced, non-axisymmetric, low frequency oscillation modes 
whose toroidal component
of the displacement vector is dominant over the horizontal and the radial components, and
whose oscillation frequency
is proportional to the rotation frequency $\Omega$
(e.g., Papaloizou \& Pringle 1978; Provost, Berthomieu, \& Rocca 1981; Saio 1982; 
Berthomieu \& Provost 1983).
$R$ modes are retrograde modes in the corotating frame of the star, but are
prograde when seen in an inertial frame.
In the limit of $\Omega\rightarrow 0$, the displacement vector is well represented by a single
spherical harmonic function $Y_{l^\prime}^m$ and the oscillation frequency $\omega$
observed in the corotating frame of the star tends to 
the asymptotic frequency given by $\omega_c(m,l^\prime)\equiv 2m\Omega/[l^\prime(l^\prime+1)]$
with $l^\prime\ge|m|$.
$R$ modes are therefore conveniently classified by using the two integer indices 
$m$ and $l^\prime(\ge|m|)$ that define the asymptotic freqyency $\omega_c(m,l^\prime)$.
For a neutrally-stratified (isentropic) star, in which the Brunt-V\"ais\"al\"a frequency $N$ 
vanishes identically in the interior,  
the odd $r$ mode with $l^\prime=m$ that has no radial nodes of the eigenfunction is
the only $r$ mode we can find for given $m$ and $\Omega$
(see, e.g, Saio 1982, Lockitch \& Friedman 1999, Yoshida \& Lee 2000a).
The restoring force for the $r$ mode is exclusively attributable to the Coriolis force.
For a stably-stratified star, however, buoyant force comes into play 
as an additional restoring force for the low frequency oscillations.
For any combination of $m$, $l^\prime(\ge|m|)$, and $\Omega$, 
there exist an infinite sequence of $r$ modes that differ in the number of radial nodes
of the eigenfunction (e.g., Yoshida \& Lee 2000b).
In this paper, we shall call $r$ modes in stably-stratified stars 
buoyant $r$ modes.

For $r$ modes to be pulsationally excited in SPB stars, rapid rotation of the star is 
required (Savonije 2005, Townsend 2005b).
Expecting a fraction of SPB stars are rapidly rotating, and noting 
that some Be stars are found in the SPB domain in the HR diagram, it may be reasonable to expect 
that there exists a link between a class of Be stars and rapidly rotating SPB stars.
Be stars are characterized by emission lines, which are thought to have their origin in the
gaseous discs surrouonding the central B stars. 
Time variability of the emission lines in strength and shape as well as their
appearance and dissapearance are believed to be
closely related to the dynamical status of the gaseous discs aurrounding the stars
(see a recent review by Porter \& Rivinius 2003).
Although discs around Be stars are indispensable for observed Be phenomena, no definite physical
mechanisms responsible for disc formation have not so far been indentified.
Considering that both $g$ modes and $r$ modes become pulsationally unstable in rapidly
rotating SPB stars, it is tempting to speculate these low frequency oscillation modes
play a role in angular momentum transfer in the interior that leads to
disc formation around the stars.
To investigate angular momentum redistribution in rotating stars, 
in a series of papers Ando (1981, 1982, 1983, 1986)
has applied to stellar pulsations 
the wave-meanflow intercation formalism, which has been extensively used
in geophysics and fluid mechanics (e.g., Lighthill 1978; Craik 1985; Pedlosky 1987;
Andrews, Holton, \& Leovy 1987).
Carring out a numerical simulation,
Ando (1986) has argued that quasi-periodic vacillation phenomena in Be stars occur
as a result of the interaction between the rotation and two prograde and retrograde global
oscillation modes.
Quite similar calculation has also been done to simulate the quasi-biennial oscillation
in the earth's atmosphere (Plumb 1977; see also Lindzen \& Holton 1968).
We think it worthwhile to
apply the wave-meanflow interaction formalism to pulsationally unstable low frequency modes 
in rapidly rotating SPB stars, speculating that
unstable prograde and retrograde $g$ modes are responsible for quasi-periodioc
vacillations, and unstable $r$ modes are working as a stable angular momentum supplier
to the surrounding discs.

In the analyses of nonadiabatic  
oscillations of SPB stars, Savonije (2005) and Townsend (2005a,b) exclusively used
the traditional approximation, neglecting the local horizontal component of the rotation vector
$\pmb{\Omega}$ in the momentum conservation equation (e.g., Eckart 1960).
In the traditional approximation,
separation of variables between spherical polar coordinates $(r,\theta,\phi)$ 
becomes possible and the governing equations for oscillations in rotating stars are 
largely simplified (e.g., Lee \& Saio 1987a).
The traditional approximation has been extensively employed in geophysics
to study waves propagating in an atmospheric or ocean fluid layer that is very thin
campared to the radius of the earth.
The approximation is thought to be
valid for low frequency oscillations with $\omega<<N$ so that the horizontal component
of the fluid velocity is much larger than the radial component (e.g., Eckart 1960).
When we apply the traditional approximation
to global oscillations of rotating stars, however, validity of the approximation is 
not very clear,
since the condition $\omega<<N$ may not always be satisfied in the entire interior, for example.
In this paper, we calculate nonadiabatic low $m$ buoyant $r$ modes of SPB stars 
without applying the traditional approximation, in order to examine their pulsational stability.
We also apply the wave-meanflow interaction formalism to nonadiabatic $r$ modes to see
if acceleration of the surface rotation is possible as a result of the forcing.
The result of the stability analysis is given in \S 3 and the wave-meanflow interaction
for the $r$ modes is discussed in \S 4.
A brief comparison between nonadiabatic $r$ mode calculations with and without the
traditional approximtion is given in \S 5.
We briefly describe our numerical method in \S 2, and conclusions are summarized in \S 6.

\section{Numerical method}

The numerical method of calculating nonadiabatic oscillations of 
uniformly rotating stars is the same as that given in Lee \& Saio (1987b, 1993).
We employ a linear theory of oscillations to describe small amplitude pulsations
of rotating stars. 
We assume that the background state of rotating stars is axisymmetric about the rotation axis, 
and that
the time $t$ and azimuthal angle $\phi$ dependence of perturbations to the background state
is given by $\exp[i(\sigma t+m\phi)]$, where $\sigma$ is the oscillation frequency
in an inertial frame, and
$m$ is an integer representing the azimuthal wavenumber around the rotation axis.
Under these assumptions, linearized hydrodynamic equations are given by
\begin{equation}
-\omega^2\rho\pmb{\xi}+2{\rm i}\omega\rho\pmb{\Omega}\times\pmb{\xi}=
-\nabla p^\prime-\rho^\prime\nabla\Phi-\rho\nabla\Phi^\prime,
\end{equation}
\begin{equation}
\rho^\prime+\nabla\cdot(\rho\pmb{\xi})=0,
\end{equation}
\begin{equation}
\nabla^2\Phi^\prime=4\pi G\rho^\prime,
\end{equation}
\begin{equation}
{\rm i}\omega\rho T\delta s=(\rho\epsilon-\nabla\cdot\pmb{F})^\prime,
\end{equation}
and
\begin{equation}
{\delta p\over p}=\Gamma_1{\delta\rho\over \rho}+\alpha_T\Gamma_1{\delta s\over c_p},
\end{equation}
where 
$\rho$ is the mass density, $p$ is the pressure, $T$ is the temperature,
$s$ is the specific entropy, $c_p$ is the specific heat at constant pressure,
$\Phi$ is the gravitational potential,
$\pmb{F}$ is the energy flux vector, $\epsilon$ is the nuclear energy generation rate per unit mass,
$\pmb{\Omega}$ is the vector of angular velocity of rotation, 
\begin{equation}
\omega\equiv\sigma+m\Omega
\end{equation} 
is the oscillation frequency observed in the corotating frame of the star, and
\begin{equation}
\Gamma_1=\left({\partial\ln p\over\partial\ln \rho}\right)_s, \quad 
\alpha_T=-\left({\partial\ln\rho\over\partial\ln T}\right)_p.
\end{equation}
Here, $\pmb{\xi}$ is the displacement vector, and
the quantities attached by a prime $(^\prime$) and $\delta$ represents
their Eulerian and Lagrangian perturbations, respectively.
The total energy flux is given as $\pmb{F}=\pmb{F}_{\rm conv}+\pmb{F}_{\rm rad}$,
where $\pmb{F}_{\rm conv}$ is the convective energy flux, which is calculated
using a mixing length theory of turbulent convection, and $\pmb{F}_{\rm rad}$
denotes the radiative energy flux given by
\begin{equation}
\pmb{F}_{\rm rad}=-{ac\over 3\kappa\rho}\nabla T^4,
\end{equation}
where $\kappa$ is the opacity, and $a$ and $c$ denote the radiative constant and the velocity of light,
respectively.
For the pertubation of the convective flux, we assume
$\delta(\nabla\cdot\pmb{F}_{\rm conv})=0$ for simplicity.

We apply boundary conditions both at the surface and at the center of the star.
The surface boundary conditions we use at $r=R$ are $\delta p=0$,
$\delta (L_r-4\pi r^2\sigma_{\rm SB}T^4)=0$, and the condition 
that $\Phi^\prime$ is continuous at the stellar surface, where $L_r=4\pi r^2 F_r$ is the
luminosity at $r$, and $\sigma_{\rm SB}$ is the Stefan-Boltzmann constant.
The inner boundary conditions at $r=0$, on the other hand, 
come from the regularity condition of the mechanical variables 
at the center and
from the condition that oscillations are nearly adiabatic,
$\delta s/c_p\sim0$, in the deep interior.

In this paper, we employ a spherical polar coordinate system $(r,\theta,\phi)$ whose origin is
at the center of the star and the axis of $\theta=0$ is along the rotation axis.
For a given $m$, the displacement vector $\pmb{\xi}(r,\theta,\phi,t)$ is given by
a truncated series expansion in terms of spherical harmonic functions $Y_l^m(\theta,\phi)$ 
with different $l$'s as
\begin{equation}
{\xi_r\over r}=\sum_{j=1}^{j_{\rm max}} S_{l_j}(r)Y_{l_j}^m(\theta,\phi)e^{i\sigma t},
\end{equation}
\begin{equation}
{\xi_\theta\over r}=\sum_{j=1}^{j_{\rm max}}
\left[H_{l_j}(r){\partial\over\partial \theta}Y_{l_j}^m(\theta,\phi)
+{T_{l_j^\prime}(r)\over\sin\theta}{\partial\over\partial\phi}Y^m_{l_j^\prime}(\theta,\phi)\right]
e^{i\sigma t},
\end{equation}
\begin{equation}
{\xi_\phi\over r}=\sum_{j=1}^{j_{\rm max}}
\left[{H_{l_j}(r)\over\sin\theta}{\partial\over\partial \phi}Y_{l_j}^m(\theta,\phi)
-T_{l_j^\prime}(r){\partial\over\partial\theta}Y^m_{l_j^\prime}(\theta,\phi)\right]
e^{i\sigma t},
\end{equation}
and the pressure perturbation, $p^\prime(r,\theta,\phi,t)$, for example, is given as
\begin{equation}
p^\prime=\sum_{j=1}^{j_{\rm max}} p^\prime_{l_j}(r)Y_{l_j}^m(\theta,\phi)e^{i\sigma t}
\end{equation}
where $l_j=|m|+2(j-1)$ and $l_j^\prime=l_j+1$ for even modes and
$l_j=|m|+2j-1$ and $l_j^\prime=l_j-1$ for odd modes where $j=1,~2,~\cdots,~j_{\rm max}$.
Note that the function $p^\prime$ is symmetric (antisymmetric) for even (odd) modes
about the equator of the star.
Substituting the expansions (9) to (12) into linearized 
hydrodynamic equations (1) to (5), we obtain a set of coupled linear ordinary differential
equations of a finite dimension $j_{\rm max}$ for the radial expansion functions.
With the boundary conditions at the center and at the surface of the star,
we solve the set of linear differential equations of a finite dimension as an eigenvalue problem
for the oscillation frequency $\omega$ (Lee \& Saio 1987b, 1993).
Oscillation modes with $\omega_{\rm I}={\rm Im}(\omega)<0$ are pulsationally
unstable while those with $\omega_{\rm I}>0$ stable.
Note that
we do not include terms associated with the centrifugal force, that is, setting $f=0$
in the oscillation equations given in Lee \& Saio (1987b, 1993).

If $\omega$ and $\pmb{\xi}$ are respectively the eigenfrequency and the eigenfunction
of a mode governed by the oscillation equations (1) to (5), 
multiplying equation (1) by $\pmb{\xi}^*$, which is
the complex conjugate of the displacement vector $\pmb{\xi}$, 
and integrating over the whole volume of the star,
we obtain
\begin{equation}
\omega^2E+2\omega F-G=0,
\end{equation}
where
\begin{equation}
E=\int\rho\pmb{\xi}^*\cdot\pmb{\xi}dV, 
\end{equation}
\begin{equation}
F=-{\rm i}\int\rho\pmb{\xi}^*\cdot\left(\pmb{\Omega}\times\pmb{\xi}\right) dV, 
\end{equation}
\begin{equation}
G=\int \pmb{\xi}^*\cdot\left(\nabla p^\prime+\rho^\prime\nabla \Phi+\rho\nabla\Phi^\prime\right)dV.
\end{equation}
Note that $E$ and $F$ are real quantities, because 
$(\pmb{\xi}^*\cdot\pmb{\xi})^*=\pmb{\xi}^*\cdot\pmb{\xi}$ and
$\left[{\rm i}\pmb{\xi}^*\cdot(\pmb{\Omega}\times\pmb{\xi})\right]^*=
{\rm i}\pmb{\xi}^*\cdot(\pmb{\Omega}\times\pmb{\xi})$.
Since the term $G$ can be rewritten as
\begin{eqnarray}
G=\int dV\left[{1\over\Gamma_1}{\delta p^*\over p}\delta p
-\left({\delta\rho^*\over\rho}\pmb{\xi}\cdot\nabla p
+{\delta\rho\over\rho}\pmb{\xi}^*\cdot\nabla p\right)
+{(\pmb{\xi}\cdot\nabla\rho)(\pmb{\xi}^*\cdot\nabla p)\over\rho}
-{\nabla\Phi^\prime\cdot\nabla\Phi^{\prime *}\over 4\pi G}
-\alpha_T\delta p{\delta s^*\over c_p}
\right]  \nonumber\\
+\int d\pmb{S}\cdot\left(\pmb{\xi}^*p^\prime+\rho\pmb{\xi}^*\Phi^\prime
+\Phi^\prime{\nabla\Phi^{\prime *}\over 4\pi G}\right),
\end{eqnarray}
substituting $\omega=\omega_{\rm R}+{\rm i}\omega_{\rm I}$ into equation (13), we obtain
\begin{equation}
\omega_{\rm I}={1\over 2(\omega_{\rm R}E+F)}
{\int {\rm Im}\left(-\alpha_T\delta p{\delta s^*\over c_p}\right)dV},
\end{equation}
where we have assumed that $\nabla p$ is parallel to $\nabla\rho$ and that
the surface terms can be ignored because of the outer boundary conditions employed in this paper.
For oscillations of rotating stars, we may define a work integral as
\begin{equation}
w(r)=-{1\over 2(\omega_{\rm R}E+F)}
{\int_0^r {\rm Im}\left(-\alpha_T\delta p{\delta s^*\over c_p}\right)4\pi r^2dr},
\end{equation}
so that we have $w(R)>0$ for unstable modes having $\omega_{\rm I}<0$.
We can use $w(R)$ as a check on numerical consistency in nonadiabatidc calculations.
As $\Omega\rightarrow 0$, $F\rightarrow 0$, and the work integral (19) reduces to 
that for non-rotating stars.
We note that the region in which $dw(r)/dr>0$ is a driving region for the 
oscillation while the region of $dw(r)/dr<0$ a damping region.

\section{Modal property of $r$ modes in SPB stars}

Main sequence models used for pulsation calculation are computed with a standard
stellar evolution code, into which
we incorporate the opacity tables computed by Iglesias, Rogers, \& Wilson (1992), and
Iglesias and Rogers (1996).
Figure 1 shows evolutionary tracks, from the ZAMS to early hydrogen 
shell burning stages, of stars having
$M=3M_\odot$ to $M=8M_\odot$ for $X=0.7$ and $Z=0.02$, where the circles are
SPB stars tabulated in Waelkens et al (1998) and the filled circles indicate 
rapidly rotating SPB stars given in Aerts et al (1999).

We classify buoyant $r$ modes using two integer indices
$m$ and $l^\prime(\ge |m|)$ that define the asymptotic frequency 
$\omega_{c}(m,l^\prime)\equiv 2m\Omega/[l^\prime(l^\prime+1)]$ for a given $\Omega$.
Note that
the mixed modes discussed by Townsend (2005b) correspond to the odd $r$ modes with $l^\prime=m$
in our classification.
For given $(m,l^\prime)$, we introduce an integer index $n$ to indicate the radial order of
the mode, and we write $r_n$ and/or $\omega_n$ to denote the $n$ th radial order $r$ mode.
The integer $n$ usually corresponds to the number of radial nodes of the eigenfunction.
In this paper we exclusively assume $m>0$.
Then in our convention, the $r$ mode frequency
$\omega$ observed in the corotating frame is positive, 
and the frequency $\sigma$ in an inertial frame is negative. 
For given $(m,l^\prime)$ and $\Omega$, the asymptotic frequency $\omega_c$ 
also sets the upper limit to the $r$ mode frequencies in the sense that 
$\omega_c(m,l^\prime)>\omega_0>\omega_1>\omega_2>\cdots$ (but see also Yoshida \& Lee 2000a).
We may call $r_0$ with $\omega_0$ the fundamental $r$ mode.
$R$ modes with $\omega_n\cong\omega_c$ have almost solenoidal displacement vector, so that
$\nabla\cdot\pmb{\xi}=-\delta\rho/\rho\sim0$ and hence $\delta p/p\sim0$ in the interior.

\subsection{Pulsational stability of buoyant $r$ modes}

We calculate nonadiabatic low $m$ $r$ modes of uniformly rotating stars 
in main-sequence stages for $M=3M_\odot$ to $8M_\odot$ for $X=0.7$ and $Z=0.02$.
As a typical example of SPB stars, we mainly discuss 
the case for the $5M_\odot$ main-sequence models.
The results for $odd$ $r$ modes with $l^\prime=m=1$ and 2 for $\bar\Omega=\Omega/\sqrt{GM/R^3}=0.4$
are given in Figures 2 and 3, respectively, 
where the dimensionless oscillation frequency 
$\bar\omega_{\rm R}\equiv\omega_{\rm R}/\sqrt{GM/R^3}$ is
plotted as a function of ${\rm Log}T_{\rm eff}$ with $T_{\rm eff}$ being the effective temperature
from the ZAMS to the TAMS.
Here, the small thin and large thick circles indicate
pulsationally stable and unstable modes, respectively.
Note that the dimensionless limiting frequencies $\bar\omega_c(m,l^\prime)$ 
for the $r$ modes with $l^\prime=m=1$ and 2 
are respectively $\bar\omega_c=0.4$ and $0.2666$ for $\bar\Omega=0.4$.
Figure 4 is the same as Figure 2 but for $\bar\Omega=0.2$, for which $\bar\omega_c=0.2$.
Since the oscillation frequency $\omega$ of $r$ modes is essentially
proportional to $\Omega$, the angular rotation frequency as large as
$\bar\Omega\gtsim 0.1$
is necessary for the $r$ modes to be effectively excited by the opacity bump mechanism.
As shown by these figures, only high radial order $r_n$ modes having $n>>1$ 
become pulsationally unstable.
As the star evolves from the ZAMS to the TAMS,
the frequency range and the number of unstable $r$ modes of the $5M_\odot$ star gradually increases 
and the frequency spectrum becomes denser, reflecting the developement of the $\mu$
(mean-molecular-weight) gradient zone just outside the convective core.
Note that all the $r$ modes are stabilized before the star goes into 
shell hydrogen burning stages.
Compared with the case of $\bar\Omega=0.2$, the width of $T_{\rm eff}$ for unstable $r$ modes 
is wider for $\bar\Omega=0.4$ in the HR diagram, suggesting
that rapid rotation favors the excitation of $r$ modes.

As found from Figures 2 to 4, in the course of evolution from ZAMS,
the fundamental $r_0$ modes with $l^\prime=m$, which behave like 
an interface mode
having the maximum amplitude at the interface between the convective core and the radiative envelope, 
become unstable because of the $\epsilon$ mechanism (see also Townsend 2005b).
The growth rates $\eta\equiv -\omega_{\rm I}/\omega_{\rm R}$ of the modes, however, 
are of order of $10^{-12}$ to $10^{-11}$, 
and the unstable fundamental $r_0$ modes may not have important observational consequences.

In Figure 5, we plot, as an example, the work integral $w(r)$ for a unstable odd $r_{20}$ mode
with $l^\prime=m=1$ of a $5M_\odot$ model with ${\rm Log} T_{\rm eff}=4.188$ for 
$\bar\Omega=0.4$.
In this figure, we also plot
$\kappa_{\rm ad}=(\partial\ln\kappa/\partial\ln p)_{\rm ad}$ versus $r/R$
for the model.
Strong excitation for the $r_{20}$ mode occurs in an outer envelope region
where $d\kappa/dr$ is positive, and 
overcomes damping effects in the deep interior.
In Figure 6, the growth rate $\eta$ of unstable
$r$ modes with $l^\prime=m=1$, 2, and 3 of the $5M_\odot$ model with ${\rm Log} T_{\rm eff}=4.188$
is plotted versus
the inertial frame oscillation frequency $|\sigma|$ for $\bar\Omega=0.4$ (0.0118mHz), 
where the two filled wedges on the horizontal
axis indicates the rotation frequencies $\Omega$ and $2\Omega$.
The $r$ modes with $l^\prime=m=1$ have on average larger growth rates, 
compared to those of the $r$ modes with $l^\prime=m=2$ and 3.

In Figure 7, we plot $\bar\omega_{\rm R}$ 
of $even$ $r$ modes with $l^\prime=m+1=2$
versus ${\rm Log}T_{\rm eff}$ for $\bar\Omega=0.4$.
No pulsationally unstable $r$ modes with even parity are found.
This may be attributable to the fact that the even $r$ modes with $l^\prime=m+1=2$ 
for a given $\Omega$ have frequencies by about a factor of 3 smaller than those of the odd $r$ modes
with $l^\prime=m=1$.
In fact, we have tried to find unstable even $r$ modes assuming $\bar\Omega=0.8$, but failed. 
Figure 8 depicts the work integral $w(r)$ of an even $r_{20}$ mode with $l^\prime=m+1=2$
of the $5M_\odot$ model with ${\rm Log} T_{\rm eff}=4.188$ for $\bar\Omega=0.4$.
The damping effect in the deep interior is by a factor of 2 to 3 larger than the excitation
near the surface. 

It may be instructive to mention a result for low $m$ $r$ modes of more massive stars,
since the instability for low frequency oscillation modes weakens as the stellar mass increases.
Figure 9 shows $\bar\omega_{\rm R}$ of odd $r$ modes with $l^\prime=m=1$ of
$8M_\odot$ main-sequence models from the ZAMS to the TAMS
for $\bar\Omega=0.6$, where the small thin and large thick circles indicate
pulsationally stable and unstable modes, respectively.
For $\bar\Omega=0.6$, high radial order
$r$ modes with $l^\prime=m=2$ and 3 become unstable too in the course of evolution,
although only the odd $r$ modes with $l^\prime=m=1$ become unstable for $\bar\Omega=0.4$.

\subsection{Eigenfunctions of buoyant $r$ modes}

Let us first show some examples of the eigenfunctions (expansion functions) 
of low $m$ $r$ modes of the $5M_\odot$ ZAMS star, which
consists of the convective core and the radiative envelope that contains
two thin convection zones in the outer part of the envelope.
Figure 10 depicts the eigenfunction $x {\rm Re}(iT_{l^\prime_1})$
of $odd$ $r$ modes with $l^\prime=m=1$ (panel a) and
of $even$ $r$ modes with $l^\prime=m+1=2$ (panel b) for $\bar\Omega=0.4$, 
where $x=r/R$, and the amplitude
normalization is given by ${\rm Re}(S_{l_1})=1$ at the surface.
Note that in the vicinity of the outer boundary of the convective core,
the even $r$ modes have a node, which is not necessarily apparent in the figure.
Since $\nabla-\nabla_{\rm ad}$ is assumed to vanish in the convective core,
$iT_{l^\prime_1}$ is almost constant there (see, e.g., Saio 1982).
Although the amplitude of the $r_0$ mode has the maximum just outside the
convective core, the amplitude maximum shifts to the surface,
as the radial order $n$ and the number of radial nodes
in the radiative envelope increase.
It is important to note that for the even $r$ modes with $l^\prime=m+1=2$,
the isentropic convective core is an evanescent region for the oscillation and 
no fundamental mode that has no radial nodes
of the eigenfunction ${\rm Re}(iT_{l^\prime_1})$ exists.
We find this is also the case for both even and odd $r$ modes if $l^\prime>|m|$.

Figure 11 shows the eigenfunction $x {\rm Re}(iT_{l^\prime_1})$
of odd $r$ modes with $l^\prime=m=1$ of a $5M_\odot$ model with
${\rm Log}T_{\rm eff}=4.188$ for $\bar\Omega=0.4$, where
the long dash-dotted line, short dash-dotted line, dotted line, and solid line
indicate $r_0$, $r_1$, $r_{10}$ and $r_{20}$ modes, respectively, and the amplitude
normalization is given by ${\rm Re}(S_{l_1})=1$ at the surface.
This model has the $\mu$ gradient zone just outside 
the convective core.
The fundamental $r_0$ mode, which is shown in the inlet, behaves like an interface mode and 
the amplitude has the sharp maximum at the interface between the convective core and 
the $\mu$-gradient zone.
We also note that the high radial order $r$ modes have very short radial wavelengthes
in the $\mu$-gradient zone, which leads to the dense frequency spectrum.

\section{Forcing on rotation by buoyant $r$ modes}

In the wave-meanflow interaction formalism, every physical quantity of a rotating star is decomposed
into the zonally-averaged, axisymmetric part and the non-axisymmetric, residual part,
and we regard the former as the quantity in the meanflow and the latter as the one representing
waves (oscillations) (e.g., Pedlosky 1987, Andrews, Holton, \& Leovy 1987).
The forcing equation for stellar rotation may be given by
(e.g., Lee \& Saio 1993; see also Ando 1983)
\begin{equation}
{\partial\over\partial t}j(r,t)=
-{1\over 2}{1\over r^2}{\partial\over\partial r}
[r^2\left<r\sin\theta{\rm Re}\left(\bar\rho\tilde v^\prime_\phi \tilde v^{\prime *}_r
+\bar v_\phi\tilde\rho^\prime \tilde v^{\prime *}_r\right)\right>]
-{1\over 2}\left<{\rm Re}\left(im\tilde\rho^{\prime *}\tilde\Phi^\prime\right)\right>,
\end{equation}
where
\begin{equation}
j(r,t)=\left<r\sin\theta\bar\rho\bar v_\phi\right>, \quad
\bar{v}_\phi={1\over 2\pi}\int_0^{2\pi}{v}_\phi d\phi, \quad 
\bar{\rho}={1\over 2\pi}\int_0^{2\pi}\rho d\phi,
\end{equation}
and
\begin{equation}
\left<\cdots\right>={1\over 2}\int_0^\pi d\theta\sin\theta (\cdots),
\end{equation}
and we rewrite the forcing equation (20) as 
\begin{equation}
{1\over\tau_0(r)} 
={1\over\tau_1(r)}+{1\over\tau_2(r)}+{1\over\tau_3(r)},
\end{equation}
where $\tau$'s are time scales of local variation of the meanflow defined as
\begin{equation}
{1\over\tau_0(r)} = {1\over j(r,t)}{\partial j(r,t)\over \partial t},
\end{equation}
\begin{equation}
{1\over\tau_1(r)} = -{1\over 2}{1\over j(r,t)}{1\over r^2}{\partial\over\partial r}
[r^2\left<r\sin\theta{\rm Re}\left(\bar\rho\tilde v^\prime_\phi \tilde v^{\prime *}_r
\right)\right>], 
\end{equation}
\begin{equation}
{1\over\tau_2(r)} = -{1\over 2}{1\over j(r,t)}{1\over r^2}{\partial\over\partial r}
[r^2\left<r\sin\theta{\rm Re}
\left(\bar v_\phi\tilde\rho^\prime \tilde v^{\prime *}_r\right)\right>], 
\end{equation}
\begin{equation}
{1\over\tau_3(r)} = -{1\over 2}{1\over j(r,t)}\left<{\rm Re}\left(im\tilde\rho^{\prime *}
\tilde\Phi^\prime\right)\right>,
\end{equation}
and positive $\tau_0$ stands for acceleration of the rotation.
Note that
$\pmb{v}^\prime={\rm i}\omega\pmb{\xi}$ for uniformly rotating stars.
The importance of non-conservative (e.g., non-adiabatic) effects 
accompanied with the oscillation will be apparent by rewriting the Eulerian formulation
into a Lagrangian one (e.g, Ando 1986b; see also Andrews \& McIntyre 1978).
The time derivative of the total angular momentum, $\dot J$, may be given by
\begin{equation}
\dot J=\int_0^R4\pi r^2{\partial j\over\partial t}dr=-2\pi
R^2\left<R\sin\theta{\rm Re}\left(\bar\rho\tilde v^\prime_\phi \tilde v^{\prime *}_r
+\bar v_\phi\tilde\rho^\prime \tilde v^{\prime *}_r\right)\right>
-\int_0^R2\pi r^2\left<{\rm Re}\left(im\tilde\rho^{\prime *}\tilde\Phi^\prime\right)\right>dr,
\end{equation}
where the first term on the right hand side stands for the angular momentum luminosity,
and the second term is the gravitational torque.

In Figure 12, we plot $\tau^{-1}_j(r)$ with $j=0$, 1, and 2 versus $r/R$ 
for a unstable $l^\prime=m=1$ $r_{20}$ mode of the $5M_\odot$ model
with ${\rm Log}T_{\rm eff}=4.188$ for $\bar\Omega=0.4$, where the solid line, dotted line,
and dashed lines represent $\tau^{-1}_0(r)$, $\tau^{-1}_1(r)$, and $\tau^{-1}_2(r)$, 
respectively, and the amplitude normalization for the mode is given by
${\rm Re}(S_{l_1}(R))=1$.
For self-excited oscillations, the term $\tau_3^{-1}$ is negligible compared to the other 
two terms and is not shown in the figure.
Although the term $\tau_2^{-1}(r)$ was ignored in the analysis by Ando (1983),
the term for low frequency oscillation modes can be more important than 
the term $\tau_1^{-1}$ in the surface region because of
the boundary condition $\delta p=0$ we apply at the surface.
Since $\tau_0^{-1}(r)$ is positive in the region of $r\sim R$,
the unstable $r_{20}$ mode contributes to accelerating the surface rotation.
This acceleration of the surface rotation is brought about by many unstable 
$l^\prime=m$ $r$ modes in SPB stars as shown by Figure 13, in which
$\tau_0^{-1}(R)$ and $\dot J/10^{34}$ for 
odd $r$ modes with $l^\prime=m=1$ of the $5M_\odot$ model
for $\bar\Omega=0.4$ are plotted versus
the dimensionless oscillation frequency $\bar\omega_{\rm R}$.
Here, the circles and squares stand for
$\tau_0^{-1}(R)$ and $\dot J/10^{34}$, respectively, and the filled and open symbols indicate
pulsationally unstable and stable modes, respectively.
Note that to calculate the quantities $\tau_0^{-1}(R)$ and $\dot J$, 
we have employed the amplitude normalization given by
${\rm Re}(S_{l_1}(R))=1$ for the modes.
We find that $\tau_0^{-1}(R)$ is positive and $\dot J$ is negative
for the pulsationally unstable $l^\prime=m=1$ $r$ modes of the model.
Negative $\dot J$ means a loss of angular momentum from the star, or equivalently
a positive angular momentum luminosity through the surface of the star.

The numerical values of $\dot J$ in Figure 13 can be compared with
observationally estimated ones.
The amount of angular momentum supply in a unit time necessary to sustain
a decretion disc around the central star in a Be system may be estimated as
(e.g., Lee, Saio, \& Osaki 1991)
\begin{equation}
\dot J_D=\dot M_D\sqrt{GMR}=1.9\times10^{34}{\dot M_D\over
10^{-10}M_\odot/yr}\left({M\over M_\odot}\right)^{1/2}\left({R\over R_\odot}\right)^{1/2},
\end{equation}
where $\dot M_D$ denotes the mass decretion rate through the disc, and $M$ and $R$
are the mass and radius of the central star, and an estimation of 
$\dot M_D\sim 10^{-10}M_\odot/yr$ given by Okazaki (2001) has been used.
Since the quantity $\dot J$ defined in equation (28) is proportional to
the sqaure of the oscillation amplitudes, if we consider the typical value of $\dot J_D$ is
of order of $10^{35}$ to $10^{36}$ for a decretion disc in Be stars, the oscillation
amplitudes of order of ${\rm Re}(S_{l_1}(R))=0.01\sim 0.1$ may be able to produce $\dot J$
comparable to the observational estimations $\dot J_D$.

\section{Comparison with the traditional approximation}

It may be appropriate here to make a brief comparison between numerical results
obtained with and without the traditional approximation for $r$ mode 
oscillations of rotating stars.
In this section, for both calculations with and without the traditional approximation,
we consistently employ the Cowling approximation,
neglecting the Eulerian perturbation of the gravitational potential.

The traditional approximation has been used by several authors 
(e.g., Ushomirsky \& Bildsten 1998, Townsend 2005a,b; Savonije 2005)
to numerically investigate low frequency oscillations of SPB stars.
Under the traditional approximation, we ignore the local horizontal
component of the rotation vector $\pmb{\Omega}$
in the momentum conservation equation. 
Adiabatic oscillation equations in the Cowling approximation then becomes
self-ajoint, 
and separation of variables becomes possible between spherical polar coordinates
$(r,\theta,\phi)$ so that an oscillation mode of a rotating star can be represented by
\begin{equation}
p^\prime(r,\theta,\phi,t)=p^\prime_{kmn}(r)H_{km}(\cos\theta)e^{{\rm i}m\phi+{\rm i}\sigma t},
\end{equation}
where $n$ denotes the radial order, and
the Hough function $H_{km}(\cos\theta)$ with integer indices $k$ and $m$
is the solution to Laplace's tidal equation and depends on the ratio $2\Omega/\omega$
(e.g., Bildsten, Ushomirsky, \& Cutler 1996; Lee \& Saio 1997).
This separation of variables greatly simplifies the problem of analyzing numerically and
analytically low frequency oscillations of rotating stars, although couplings between oscillation 
modes associated with different Hough functions
cannot be treated under the approximation.
In geophysics, the traditional approximation has long been used to discuss
low frequency atmospheric or ocean waves that satisfy $N/\omega>>1$ and have
the horizontal velocity much larger than the vertical (radial) velocity (e.g., Eckart 1960).
Although atmospheric or ocean waves discussed in geophysics are propagating in a layer 
much thinner compared to the
radius of the earth, low frequency modes of a rotating star in general
have eigenfunctions extending from the center to the surface of the star and do not always
satisfy in the interior the conditions needed for the tradiational approximation to be valid.

A typical difficulty in applying the traditional approximation to stellar pulsations appears
in spurious singular behavior of the eigenfunctions at the center 
for certain groups of low frequency modes.
In the tradiational approximation, the governing equations for adiabatic
oscillations may be given by (e.g., Lee \& Saio 1987a)
\begin{equation}
r{dz_1\over dr}=\left({V\over\Gamma_1}-3\right)z_1+\left({\lambda_{km}\over c_1\bar\omega^2}
-{V\over\Gamma_1}\right)z_2,
\end{equation}
\begin{equation}
r{dz_2\over dr}=\left(c_1\bar\omega^2-rA\right)z_1+\left(1-rA-U\right)z_2,
\end{equation}
where the variables $z_1$ and $z_2$ respectively correspond to $\xi_r/r$ and $p^\prime/\rho gr$,
and $\lambda_{km}$ is the separation factor, which depends on the ratio
$\nu\equiv 2\Omega/\omega$ for given integer indices $m$ and $k$, and
\begin{equation}
U={d\ln M_r\over d\ln r}, \quad V=-{d\ln p\over d\ln r}, \quad
rA={d\ln\rho\over d\ln r}-{1\over\Gamma_1}{d\ln p\over d\ln r}, \quad
c_1={(r/R)^3\over M_r/M},
\end{equation}
and $M_r=\int_0^r4\pi r^2\rho dr$.
Here, we have used the notation for $\lambda_{km}$ employed by Lee \& Saio (1997).
For $k\ge 0$, $\lambda_{km}\ge0$ tends to $(|m|+k)(|m|+k+1)$
as $\nu\rightarrow 0$, while for $k<0$, we have branches of $\nu$, in which $\lambda_{km}$
is negative.
$R$ modes are associated with $\lambda_{km}$ with negative $k$ that is positive
for $\omega< 2m\Omega/[l^\prime(l^\prime+1)]$ where $l^\prime=m+|k+1|$.
Assuming both $z_1$ and $z_2$ in equations (31) and (32)
are proportional to $r^\beta$ in the vicinity of the stellar 
centre, we obtain 
the regularity condition at the centre for the functions $z_1$ and $z_2$ given by 
\begin{equation}
\beta=(-5+\sqrt{1+4\lambda_{km}})/2\equiv\tilde l-2\ge-1,
\end{equation}
which is rewritten as
\begin{equation}
\lambda_{km}=\tilde l(\tilde l+1)\ge2.
\end{equation}
As suggested by Figure 1 of Lee \& Saio (1997), 
prograde modes associated with
$k=0$ and $m=1$ and $r$ modes with $k\le -1$ except for high radial order $k=-1$ $r$ modes 
do not satisfy the condition (35), and 
they show spurious singular behavior of the eigenfunctions $z_1$ and $z_2$
at the center that does not exist without the traditional approximation.

To avoide this mode-dependent, awkward singular behavior we meet when numerically integrating
the oscillation equations (31) and (32),
Townsend (2005a,b) introduced a new set of dependent variables $q_1$ and $q_2$ defined by
\begin{equation}
z_1=q_1x^{\beta}, \quad z_2=q_2x^{\beta},
\end{equation}
to obtain
\begin{equation}
r{dq_1\over dr}=\left({V\over\Gamma_1}-3-\beta\right)q_1+\left({\lambda_{km}\over c_1\bar\omega^2}
-{V\over\Gamma_1}\right)q_2,
\end{equation}
\begin{equation}
r{dq_2\over dr}=\left(c_1\bar\omega^2-rA\right)q_1+\left(1-rA-U-\beta\right)q_2,
\end{equation}
where $x=r/R$.
Obviously,
the variables $q_1$ and $q_2$ are always regular at the stellar center, and
we have no trouble to solve numerically equations (37) and (38) to obtain
eigenfrequencies $\omega$ of low frequency modes.

Under the traditioanl approximation,
Townsend (2005a,b) numerically solved a set of coupled linear ordinary differential equations
derived by using regularized dependent variables like $q_1$ and $q_2$
for adiabatic and non-adiabatic oscillations.
Here, we calculate non-adiabatic $r$ modes
of the $5M_\odot$ ZAMS model with $Z=0.02$, following
the formulation given by Townsend (2005a), and we compare the results to those
computed without the traditional approximation.
Figure 14 shows $\bar\omega_{\rm R}$ (panel a) and the growth rate $|\eta|$ (panel b)
of odd $r$ modes with $l^\prime=m=1$ versus the radial order $n$ for $\bar\Omega=0.4$.
From panel (a), we find that 
the two calculations for $\bar\omega_{\rm R}$ are qualitatively in good agreement with each other, 
but it is also apparent that
the difference in $\bar\omega_{\rm R}$ between the two calculations for a given radial order $n$
becomes comparable
to the difference in $\bar\omega_{\rm R}$ between two consecutive radial order modes
as the radial order $n$ increases.
Panel (b), on the other hand, shows that the growth rates of unstable $r$ modes
obtained in the traditional approximation are
systematically smaller and the instability band in the radial order $n$ is narrower, compared to
those obtained without the approximation.
We also find that the fundamental $r_0$ mode is unstable in the traditional approximation,
but it is not without the approximation.
This difference in the stability of the fundamental $r$ modes
may be attributable to the singular behavior of
the eigenfunctions in the traditional approximation, since the large amplitudes
of the eigenfunctions may wrongly strengthen the instability in the convective core.

Figure 15 plots $\bar\omega_{\rm R}$ versus the radial order $n$ for $even$ $r$ modes with
$l^\prime=m+1=2$ for $\bar\Omega=0.4$, 
where the small filled and large open circles stand for the results with and without
the traditional approximation, respectively.
Although the two calculations of $\bar\omega_{\rm R}$ are in reasonable agreement, 
it is interesting to note that 
in the traditional approximation there exists 
the fundamental $r_0$ mode
with even parity, which does not exist if we lift the traditional approximation.

In Figure 16, we plot ${\rm Re}(\lambda_{km})$ versus the radial order $n$ of $r$ modes, where
the filled and open circles respectively stand for $odd$ $r$ modes with $l^\prime=m=1$ 
and $even$ $r$ modes with $l^\prime=m+1=2$ for $\bar\Omega=0.4$.
Note that the former and the latter corresponds to $(k,m)=(-1,1)$ and 
$(-2,1)$, respectively, since $l^\prime=|m|+|k+1|$ for $r$ modes.
The figure clearly shows that the low radial order $r$ modes
with $l^\prime=m=1$ and all the $r$ modes with $l^\prime=m+1=2$ do not satisfy 
the regularity condition ${\rm Re}(\lambda_{km})\ge 2$
for the functions $z_1$ and $z_2$ at the center.
As an example, Figure 17a depicts the eigenfunctions $x{\rm Re}(z_1)$ 
and $x{\rm Re}(S_{l_1})$ for the 
fundamental $r$ modes with $l^\prime=m=1$, while Figure 17b shows the eigenfunctions
for $r_{10}$ modes with $l^\prime=m=1$, 
where $\bar\Omega=0.4$, and 
the functions $z_1$ and $S_{l_1}$ are obtained with and without the traditional approximation,
and the amplitude normalization for $z_1$ and $S_{l_1}$ is given by
${\rm Re}(z_1)=1$ and ${\rm Re}(S_{l_1})=1$ at the surface.
The function $x z_1$ of the fundamental mode 
is singular at the stellar center, but $x S_{l_1}$ is not.
On the other hand, no singular behavior appears for the
high radial order $r_{10}$ modes with $l^\prime=m=1$ even in the traditional approximation.

It seems pulsational instability of $r$ modes appears weaker in the traditional approximation.
This trend can be seen more clearly in pulsational stability of low $m$ $r$ modes
of more massive stars as shown by Figure 18, in which $\bar\omega_{\rm R}$
of $l^\prime=m=1$ $r$ modes computed for $8M_\odot$ main-sequence models 
in the traditional approximtioan is plotted
versus ${\rm Log}T_{\rm eff}$ for $\bar\Omega=0.6$, where
the samll open circles and large filled circles stand for stable and unstable modes,
respectively.
Camparing Figure 18 with Figure 9, we find the instability domain
in the traditional approximation is largely shrinked and the blue edge 
of the instability strip has shifted to the lower
$T_{\rm eff}$ in the HR diagram.

\section{Conclusions}

In this paper, we have calculated nonadiabatic $r$ modes of SPB stars
with mass ranging from $M=3M_\odot$ to $M=8M_\odot$, and found that
low $m$, odd $r$ modes with $l^\prime=m$ become pulsationally unstable 
if we assume rapid rotation of the stars, confirming the numerical results
obtained in the traditional approximation by Savonije (2005) and Townsend (2005b).
No low $m$, even $r$ modes with $l^\prime=m+1$ are found pulsationally unstable,
which contradicts Savonije (2005) who found unstable even $r$ modes.
A possible reason for the contradiction may be
that Sovonije (2005) calculated forced oscillations for $r$ modes of SPB stars, 
which are not necessarily equivalent to free $r$ mode oscillations computed in this paper.
Photometric amplitudes of the variability caused by unstable
odd $r$ modes with $l^\prime=m$ will be strongly dependent on the inclination angle between
the rotation axis and the line of sight, because the surface pattern of
the temperature perturbation is antisymmetric about the equator of the star.
Applying the wave-meanflow formalism to pulsationally unstable $r$ modes in SPB stars,
we have found that the unstable $r$ modes have a contribution to accelerating 
the surface rotation of the star.
Note that it is also important to examine the mechanism for unstable $g$ modes.

We have made a brief comparison between $r$ mode calculations done
with and without the traditional approximation.
We found that the oscillation frequency $\omega_{\rm R}$ of $r$ modes obtained
in the traditional approximation are qualitatively in good agreement with those 
without the traditional approximation.
The growth rates of unstable $r$ modes in the traditional approximation are 
systematically smaller than those
obtained without the traditional approximation, that is,
the pulsational instability appears weaker in the traditional approximation.
For $r$ modes associated with $\lambda_{km}<2$, 
the eigenfunctions obtained in the traditional approximation
show spurious singularity at the center, which may affect the pulsational stability, 
particularly for the low radial order modes.
Note that we need to make a comparison for the case of $g$ modes to have
a good understanding about the validity of the traditional approximation.
For low frequency $g$ modes, mode coupling between $g$ modes associated with different 
$\lambda_{km}$'s might have significant influence on the frequency spectrum and the stability, 
as suggested by Lee (2001).
For inertial modes in an isentropic star for whcih 
the Brunt-V\"ais\"al\"a frequency $N$ vanishes identically in the interior,
we found it difficult to obtain the oscillation frequencies in general.
This is quite reasonable because the traditional approximation is considered to be
valid for low frequency modes with $N/\omega>>1$.
Interestingly enolugh, however, we do obtain the frequencies corresponding to
the fundamental $r$ modes with $l^\prime=m$
for isentropic polytropes even in the traditional approximation.
The frequency $\omega$ of the $r$ mode in the traditional approximtion
is very close to $2\Omega/(|m|+1)$, which is consistent with
previous calcualtions done without the traditional approximation (e.g., Yoshida \& Lee 2000a).

Theoretically speaking,
both $g$ modes and $r$ modes are excited in rapidly rotating SPB stars.
As shown by Aerts et al (1999), a fraction of SPB stars are rapid rotators, and 
we think it is tempting to look for a link between Be stars and raidly rotating SPB stars.
Recent discovery of a Be star that also shows SPB variability caused by
$g$ modes and $r$ modes (private communication with G. Walker and with H. Saio, 2005)
indicates such a possibility.
If low $m$ $r$ modes excited in SPB stars accelerate the surface rotation effectively
as a result of angular momentum redistribution in the interior,
this acceleration can be a candidate mechanism for disc formation around the
rapidly rotating star.
We speculate unstable $r$ modes are a steady supplier of angular momentum to the disc and
prograde and retrograde $g$ modes also excited in SPB stars may cause 
quasi-periodic pheonomena in Be stars.
Further theoretical and observational studies are warranted.

\newpage

\begin{figure}
\centering
\epsfig{file=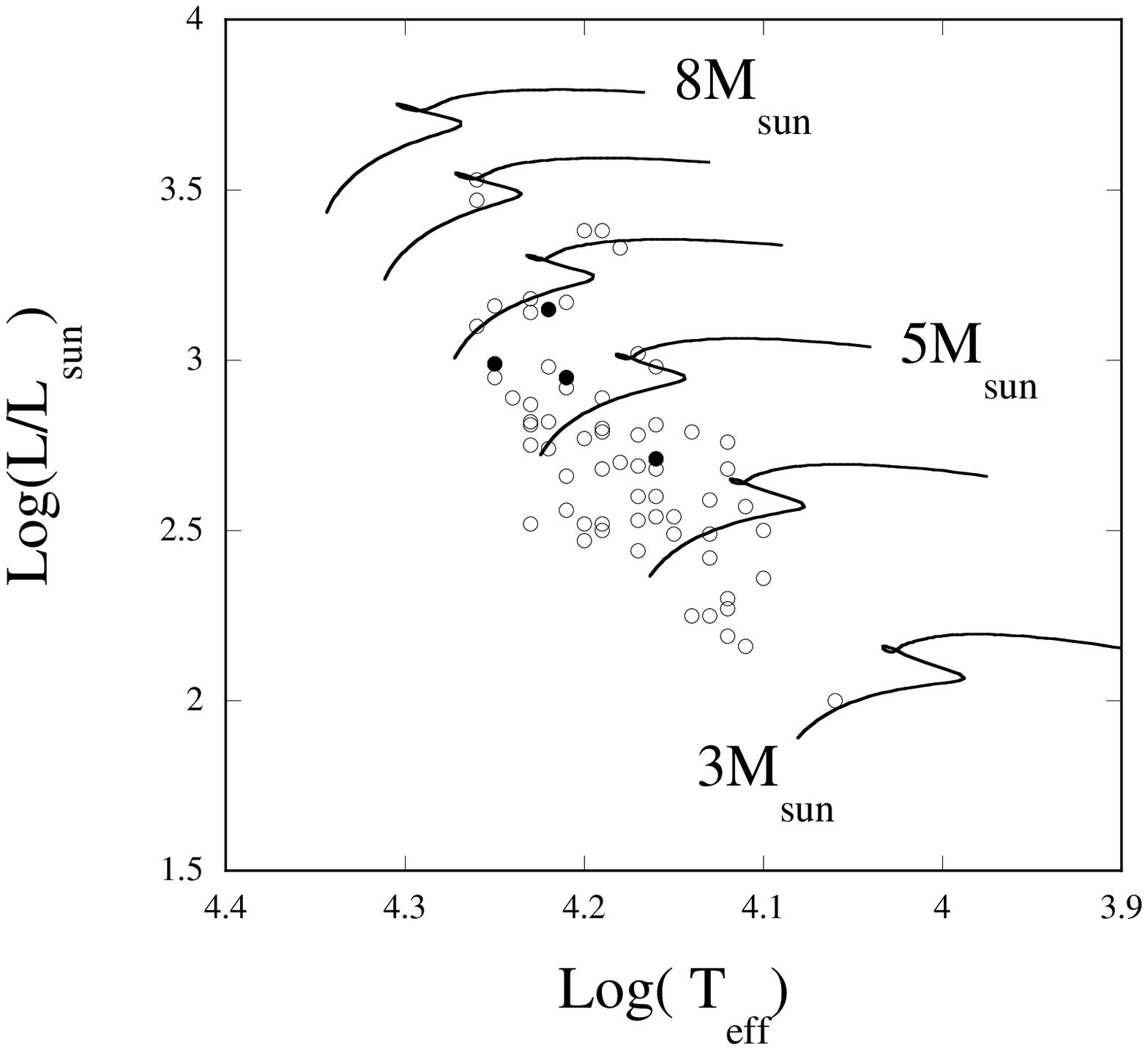,width=0.6\textwidth}
\caption{Evolution tracks of main-sequence stars with mass ranging from
$3M_\odot$ to $8M_\odot$ for $X=0.7$ and $Z=0.02$. The circles stand for SPB stars tabulated
in Waelkens et al (1998), and the filled circles indicate rapidly rotating SPB stars
as classified by Aerts et al (1999).}
\end{figure}

\begin{figure}
\centering
\epsfig{file=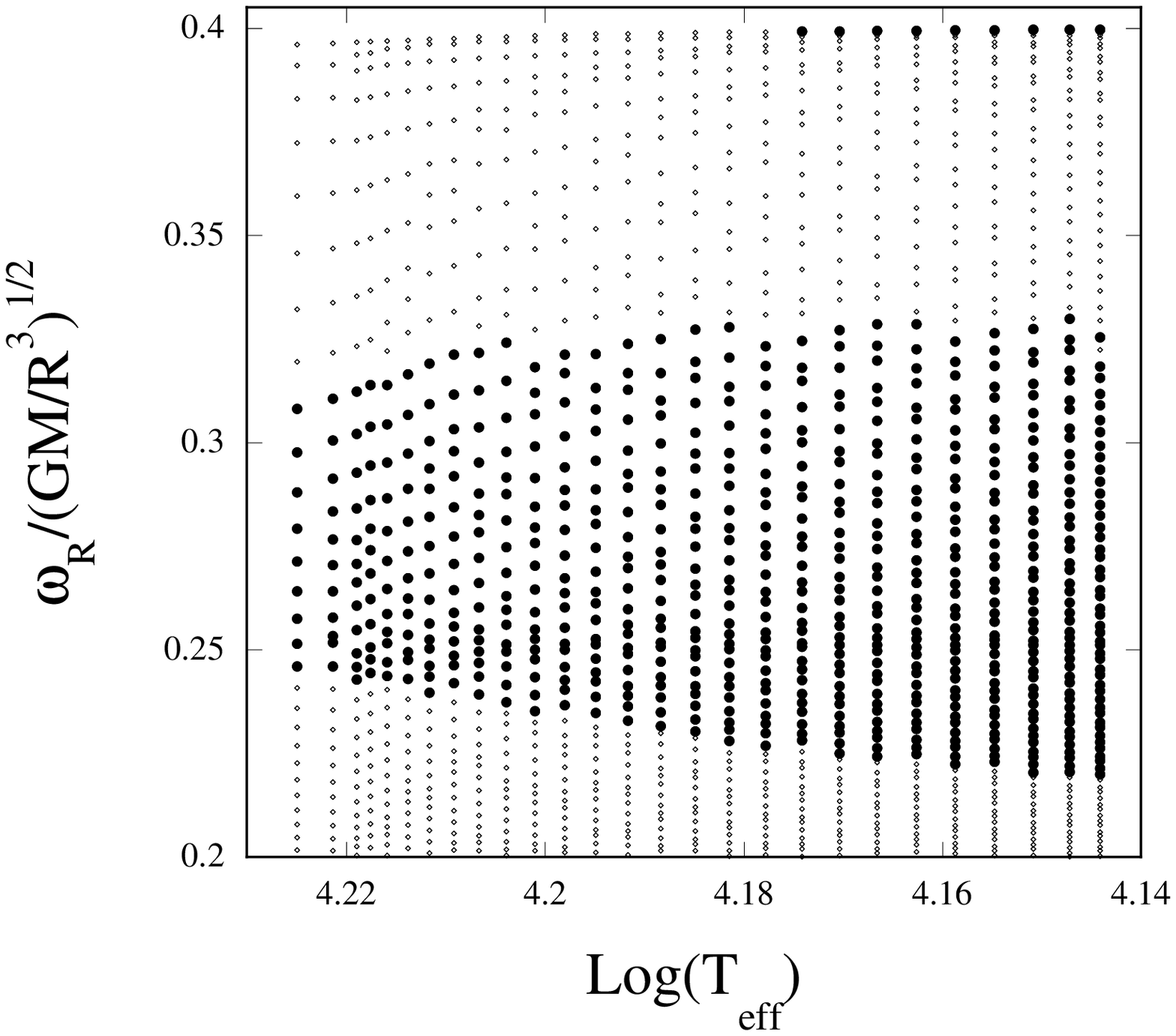,width=0.6\textwidth}
\caption{$\bar\omega_{\rm R}=\omega_{\rm R}/\sqrt{GM/R^3}$ versus 
${\rm Log}$$T_{\rm eff}$ for odd $r$ modes with $l^\prime=m=1$ of $5M_\odot$ main-sequence
models for $\bar\Omega=\Omega/\sqrt{RM/R^3}=0.4$. 
The small thin circles and large thick circles stand for pulsationally stable and unstable modes,
respectively.}
\end{figure}

\begin{figure}
\centering
\epsfig{file=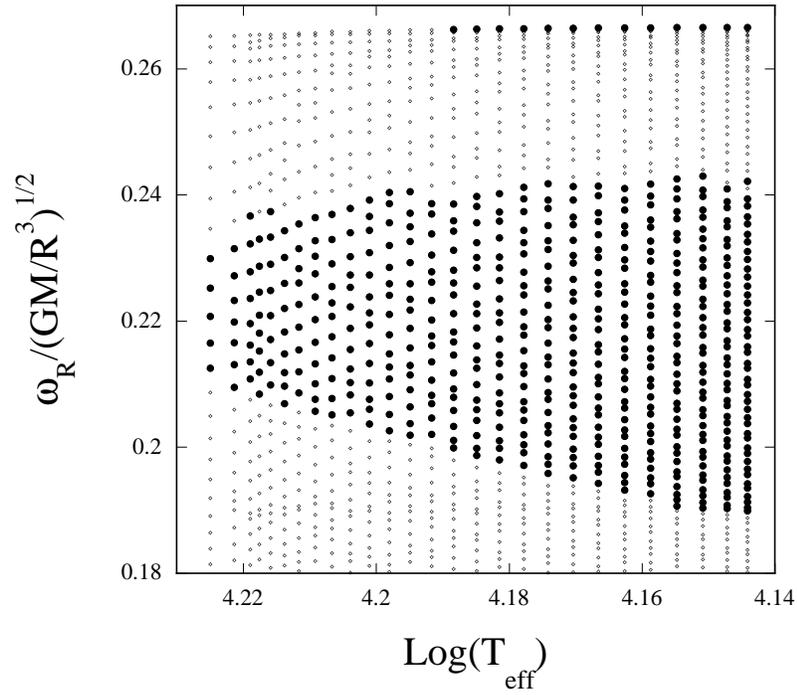,width=0.6\textwidth}
\caption{Same as Figure 2 but for odd $r$ modes with $l^\prime=m=2$. }
\end{figure}

\begin{figure}
\centering
\epsfig{file=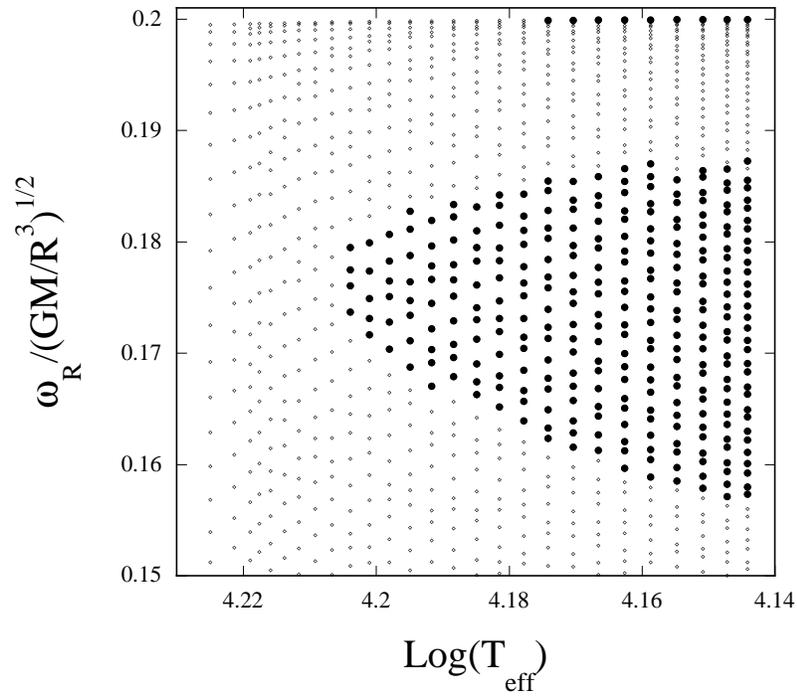,width=0.6\textwidth}
\caption{Same as Figure 2 but for $\bar\Omega=0.2$. }
\end{figure}

\begin{figure}
\centering
\epsfig{file=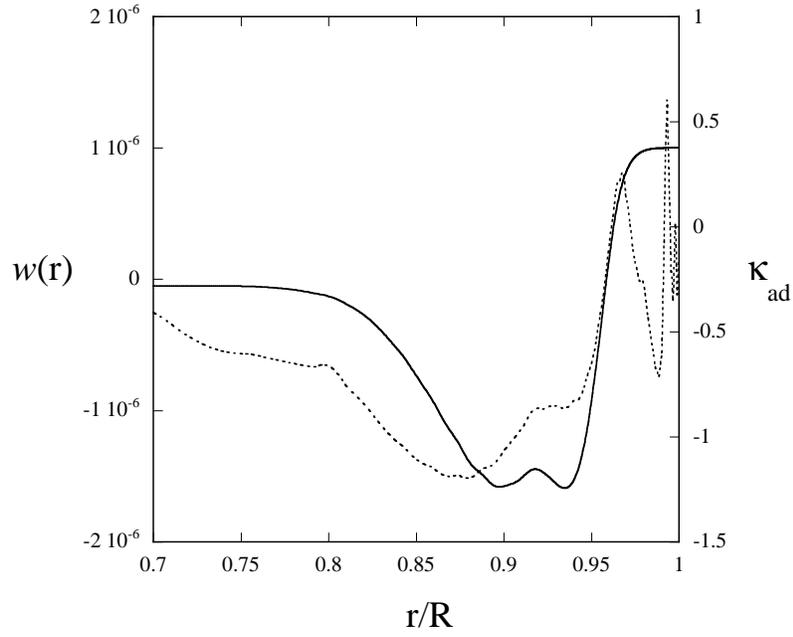,width=0.6\textwidth}
\caption{Work integral $w(r)$ of a $l^\prime=m=1$ $r_{20}$ mode and
$\kappa_{\rm ad}=(\partial\ln \kappa/\partial\ln p)_s$ for a $5M_\odot$
model with ${\rm Log}(T_{\rm eff})=4.188$, where
$\bar\Omega=0.4$ is assumed for the mode. Here,
the solid and dotted lines are for $w(r)$ and $\kappa_{\rm ad}$,
respectively.}
\end{figure}

\begin{figure}
\centering
\epsfig{file=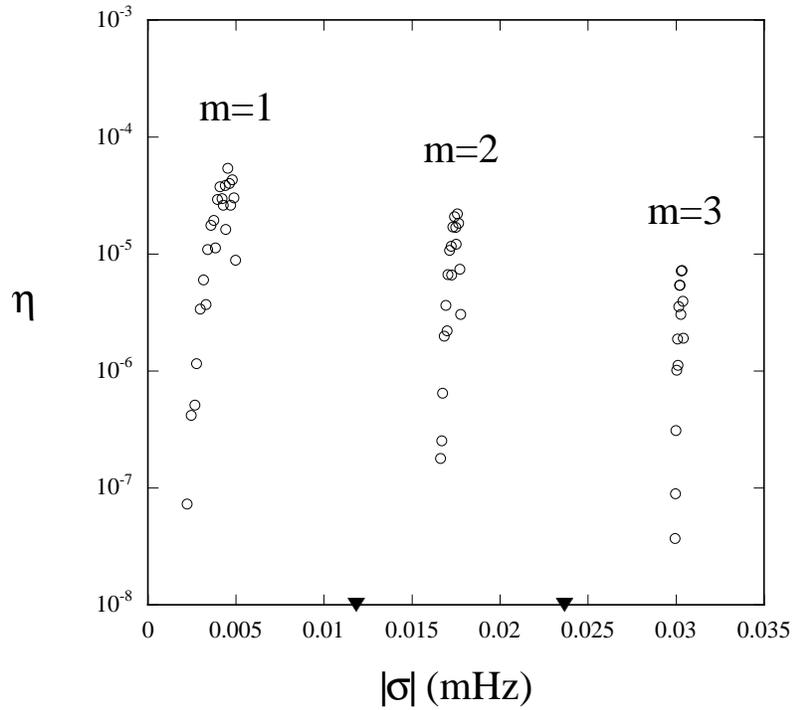,width=0.6\textwidth}
\caption{Growth rate $\eta=-\omega_{\rm I}/\omega_{\rm R}$ 
versus the inertial frame oscillation frequency
$|\sigma|$ of unstable odd $r$ modes with
$l^\prime=m=1$, 2, and 3 for a $5M_\odot$ model with ${\rm Log}(T_{\rm eff})=4.188$,
where $\bar\Omega=0.4$ (0.0118mHz). The two filled wedges on the horizontal axis
stand for $\Omega$ and $2\Omega$.}
\end{figure}

\begin{figure}
\centering
\epsfig{file=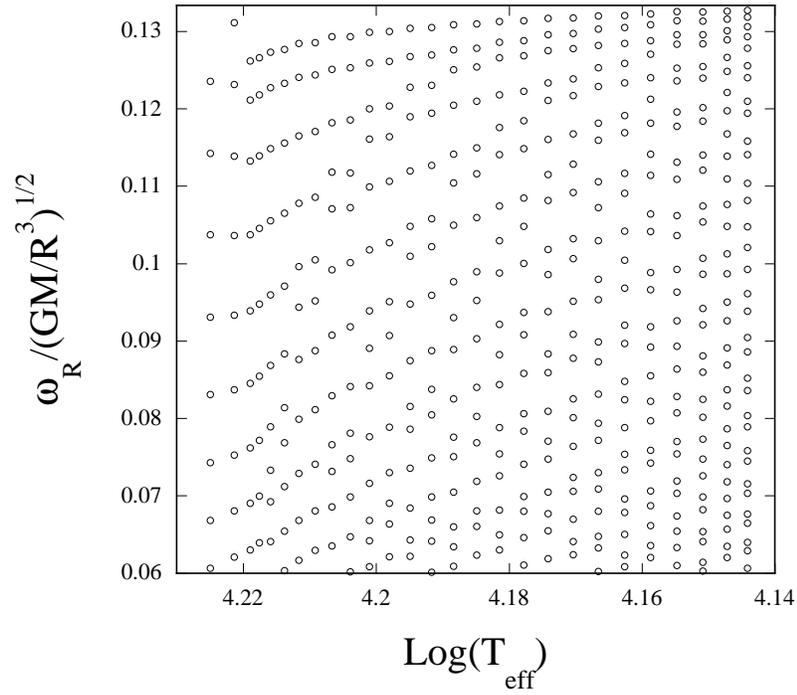,width=0.6\textwidth}
\caption{Same as Figure 2 but for even $r$ modes with $l^\prime=m+1=2$.
No even $r$ modes with $l^\prime=m+1=2$ are found pulsationally unstable.}
\end{figure}

\begin{figure}
\centering
\epsfig{file=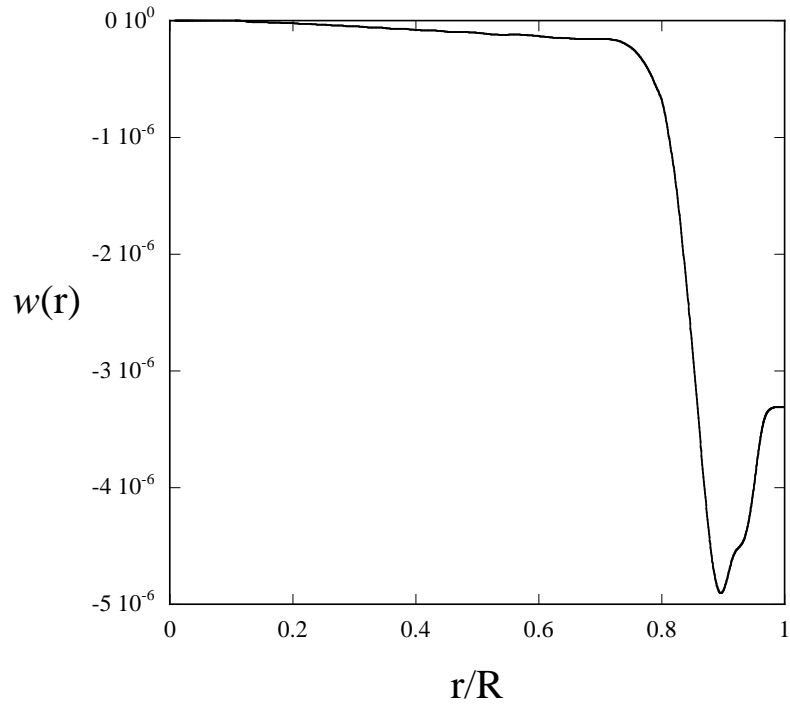,width=0.6\textwidth}
\caption{Work integral $w(r)$ of an even $l^\prime=m+1=2$ $r_{20}$ mode of a $5M_\odot$
main-sequence model with ${\rm Log}(T_{\rm eff})=4.188$, where $\bar\Omega=0.4$.}
\end{figure}

\begin{figure}
\centering
\epsfig{file=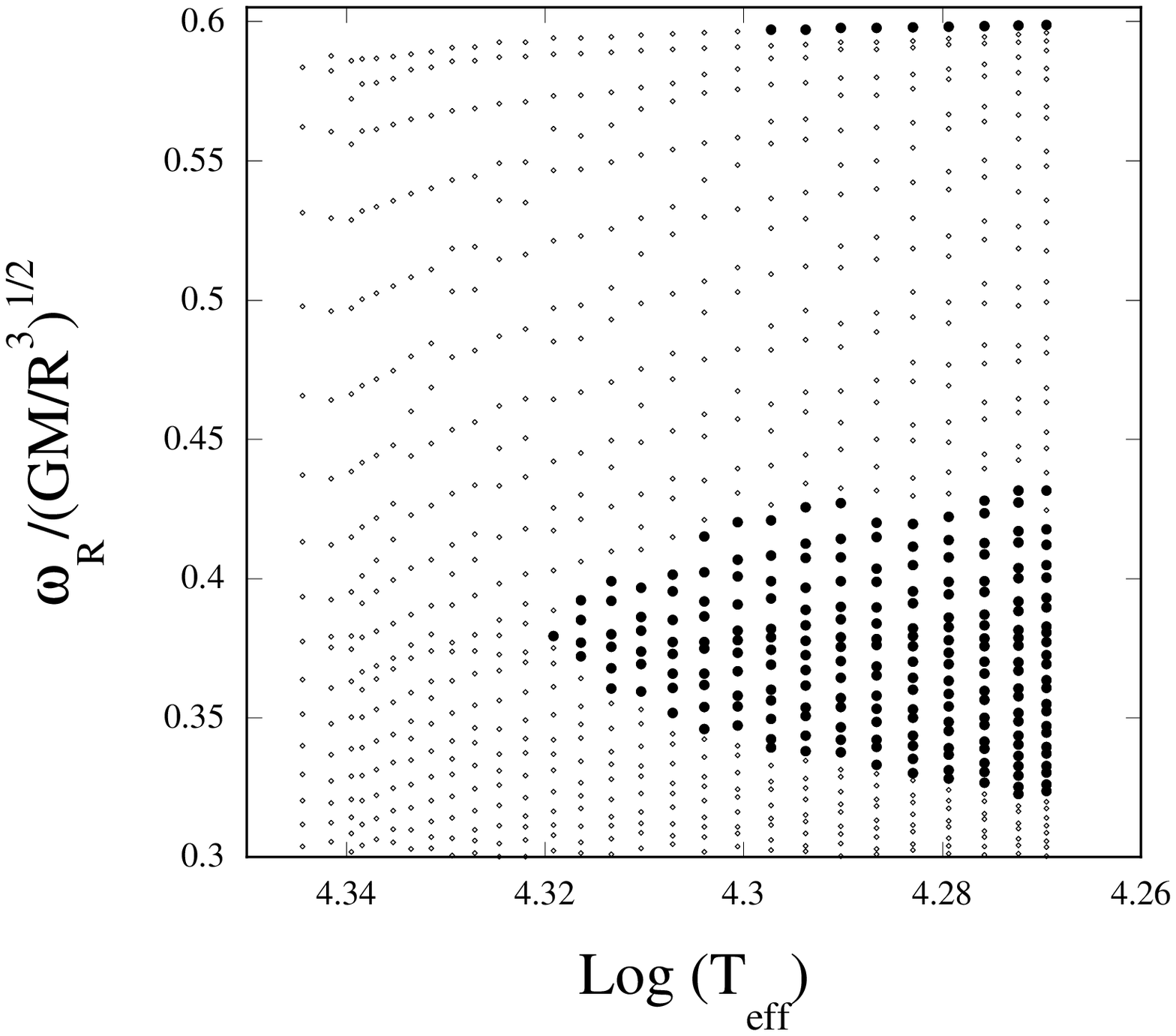,width=0.6\textwidth}
\caption{$\bar\omega_{\rm R}$ versus $\rm Log$$T_{\rm eff}$ 
for odd $r$ modes with $l^\prime=m=1$ of $8M_\odot$ main-sequence
models for $\bar\Omega=0.6$. 
The small thin circles and large thick circles stand for pulsationally stable and unstable modes,
respectively.}
\end{figure}

\begin{figure}
\centering
\epsfig{file=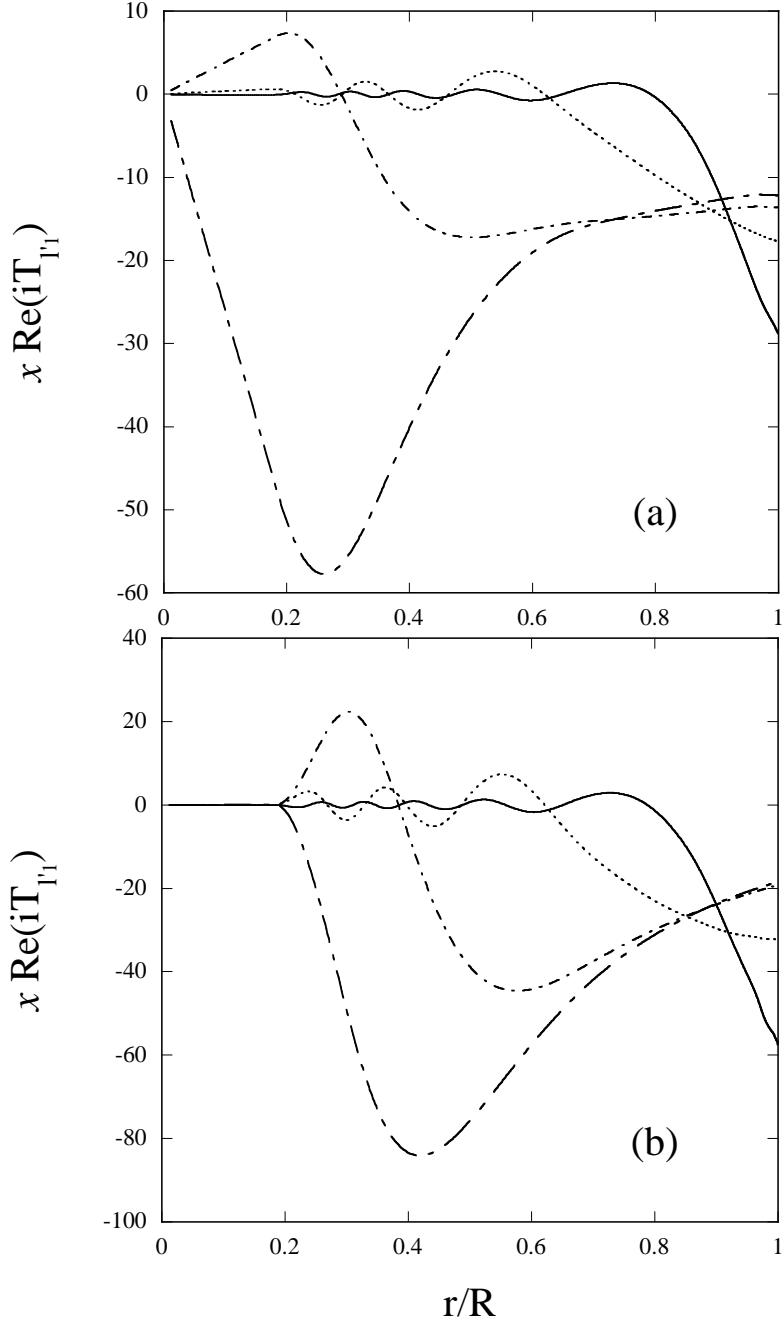,width=0.6\textwidth}
\caption{Eigenfunctions $x{\rm Re (i}T_{l'_1})$ versus the fractional radius $x=r/R$ 
for odd $l^\prime=m=1$ $r$ modes 
(panel a) and for even $l^\prime=m+1=2$ $r$ modes (panel b)
of the $5M_\odot$ ZAMS model, where $\bar\Omega=0.4$.
In panel (a), the long dash-dotted line, 
short dash-dotted line, dotted line, and solid line represent
$r_0$, $r_1$, $r_5$, and $r_{10}$ modes, respectively.
In panel (b), the long dash-dotted line, short dash-dotted line, dotted line, and 
solid line 
stand for $r_1$, $r_2$, $r_6$,and  $r_{11}$ modes, respectively.
The amplitude normalization is given by ${\rm Re}(S_{l_1})=1$ at the surface of the star.}
\end{figure}

\begin{figure}
\centering
\epsfig{file=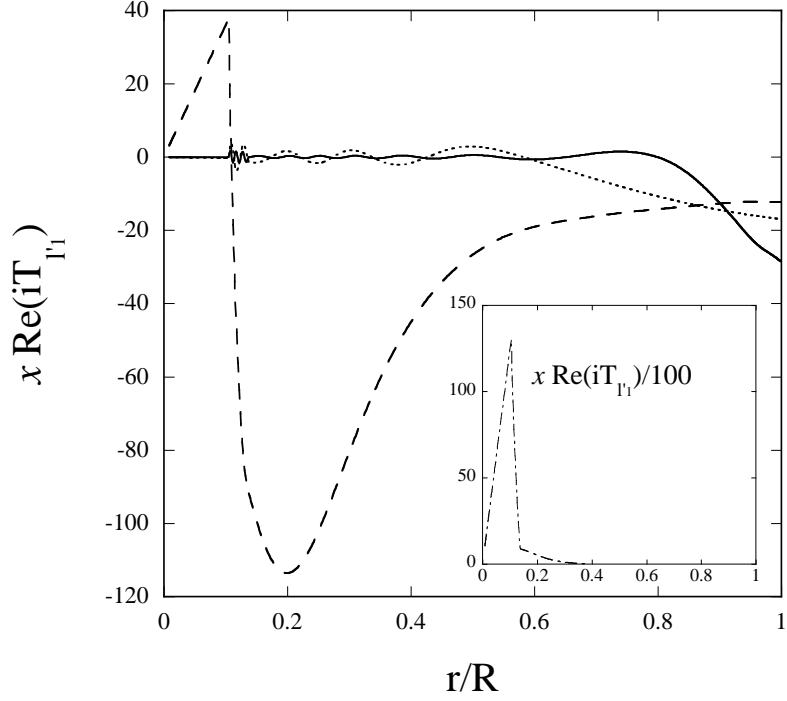,width=0.6\textwidth}
\caption{Eigenfunctions $x{\rm Re (i}T_{l'_1})$ versus $x=r/R$ 
for odd $l^\prime=m=1$ $r$ modes 
of a $5M_\odot$ model with ${\rm Log}(T_{\rm eff})=4.188$, where
$\bar\Omega=0.4$.
Here, the dash-dotted line, 
dashed line, dotted line, and solid line represent 
$r_0$, $r_1$, $r_{10}$, and $r_{20}$ modes, respectively.
The inlet shows $r_0$ mode.
The amplitude normalization is given by ${\rm Re}(S_{l_1})=1$ at the surface of the star.}
\end{figure}

\begin{figure}
\centering
\epsfig{file=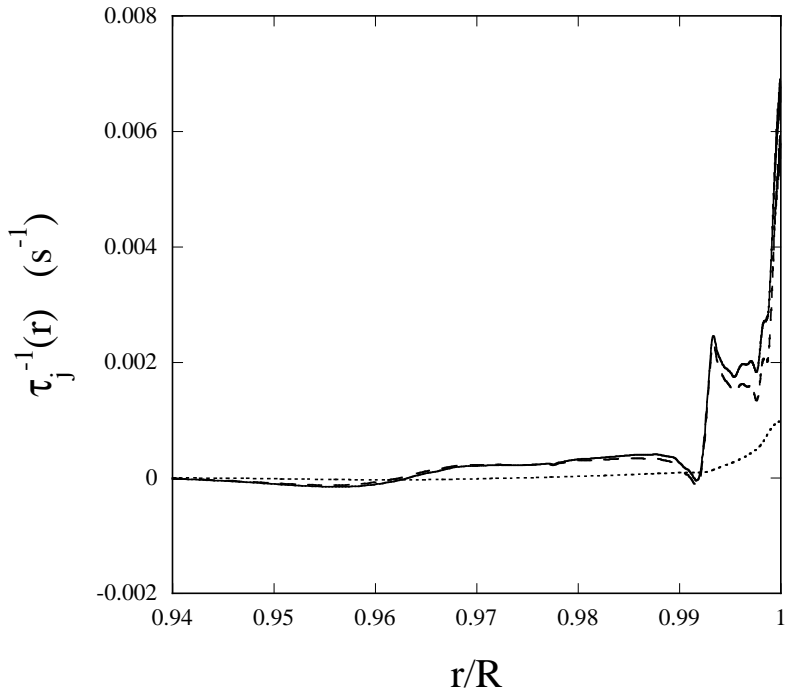,width=0.6\textwidth}
\caption{$\tau_j^{-1}(r)$ versus $r/R$ for an odd  $l^\prime=m=1$ $r_{20}$ mode 
of a $5M_\odot$ main-sequence model with ${\rm Log}(T_{\rm eff})=4.188$ for $\bar\Omega=0.4$,
where the solid line, dotted line, and dashed line stand for $\tau_0^{-1}$, $\tau_1^{-1}$,
and $\tau_2^{-1}$, respectively.
The amplitude normalization to calculate the $r_{20}$ mode
is given by ${\rm Re}(S_{l_1})=1$ at the surface of the star. }
\end{figure}

\begin{figure}
\centering
\epsfig{file=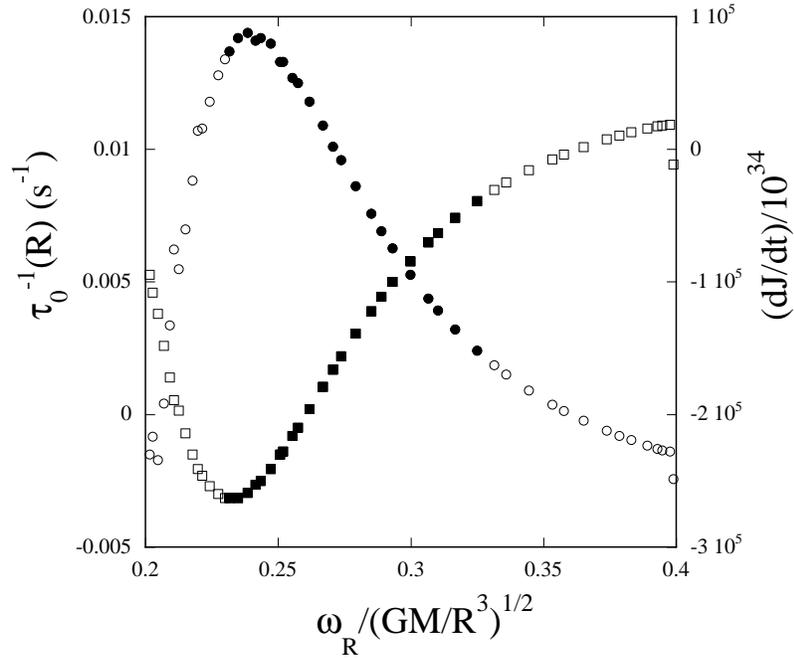,width=0.6\textwidth}
\caption{$\tau_0^{-1}(R)$ and $\dot J$ versus $\bar\omega_{\rm R}$ for
odd $r$ modes with $l^\prime=m=1$ of a $5M_\odot$ main-sequence model 
with ${\rm Log}(T_{\rm eff})=4.188$ for $\bar\Omega=0.4$,
where the circles and squares stand for $\tau_0^{-1}(R)$ and $\dot J$, respectively, and
the open and solid symbols indicate pulsationally stable and unstable modes, respectively.
The amplitude normalization to calculate the $r$ modes
is given by ${\rm Re}(S_{l_1})=1$ at the surface of the star. }
\end{figure}

\begin{figure}
\centering
\epsfig{file=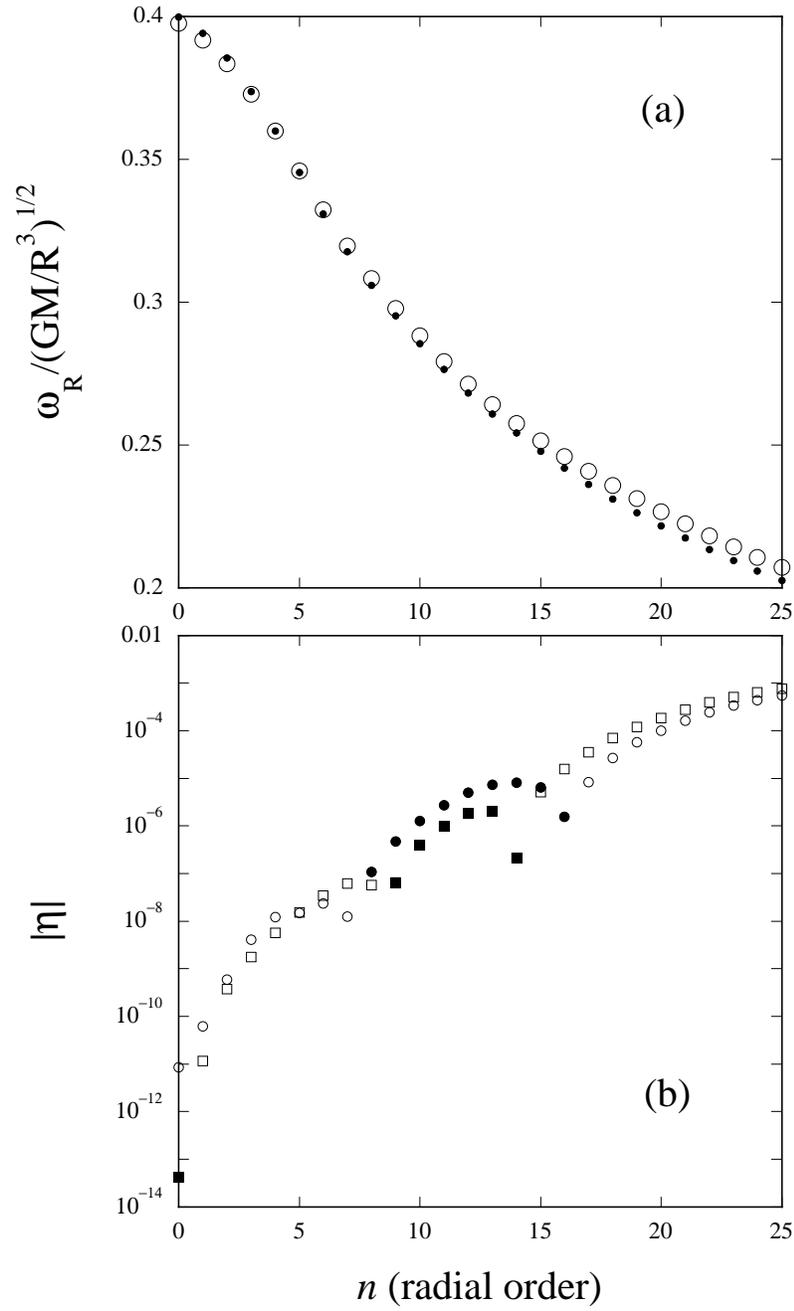,width=0.6\textwidth}
\caption{$\bar\omega_{\rm R}$ and growth rate $|\eta|$
versus the radial order $n$ of $l^\prime=m=1$ $r$ modes
of the $5M_\odot$ ZAMS model for $\bar\Omega=0.4$.
In panel (a), the small filled circles and 
large open circles stand for results obtained with and without 
the traditional approximation, respectively.
In panel (b),
the squares and circles stand for results obtained with and without 
the traditional approximation, respectively, and the filled and open symbols indicate
pulsationally unstable and stable modes, respectively.}
\end{figure}

\begin{figure}
\centering
\epsfig{file=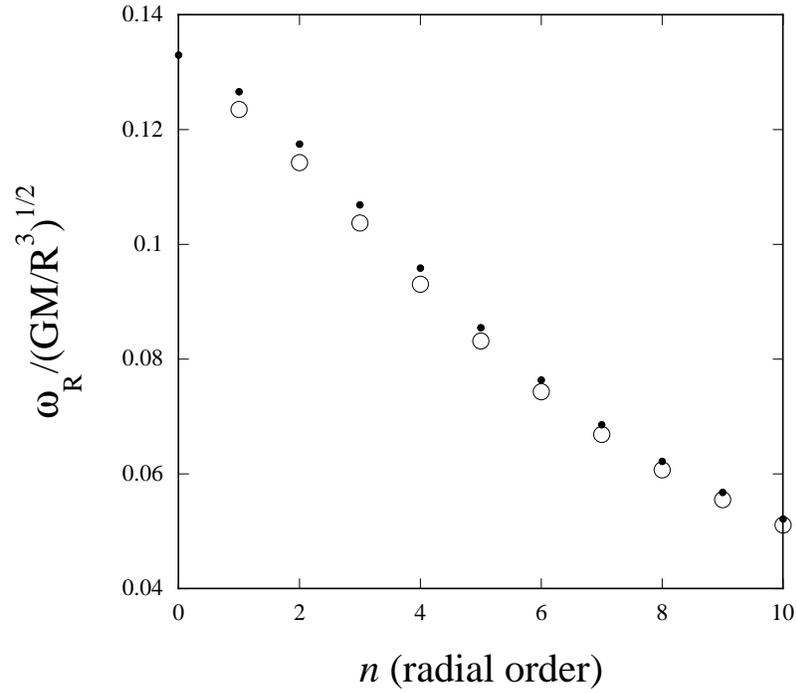,width=0.6\textwidth}
\caption{Same as Figure 14a but for even $r$ modes with $l^\prime=m+1=2$.}
\end{figure}

\begin{figure}
\centering
\epsfig{file=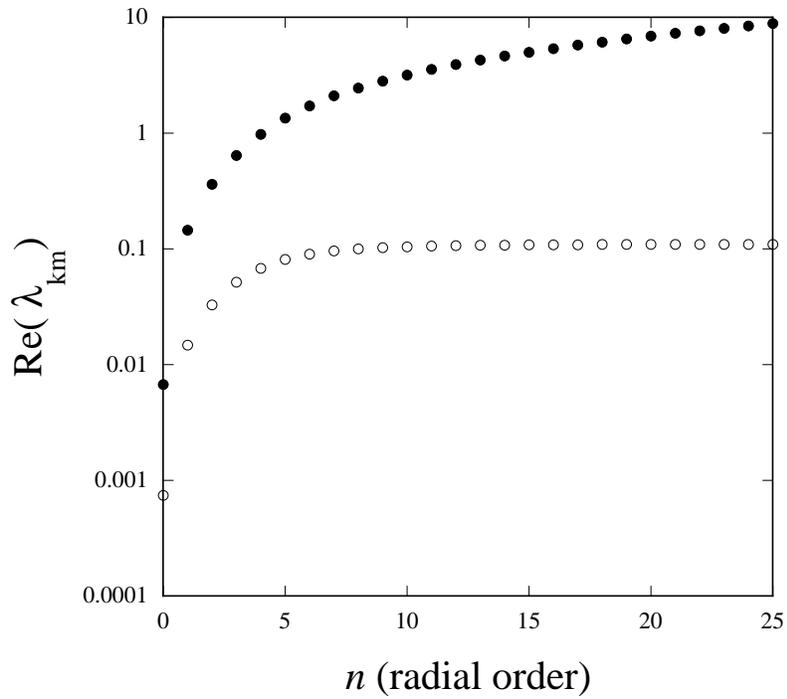,width=0.6\textwidth}
\caption{Separation factor $\lambda_{km}$ versus the radial order $n$ of $r$ modes with
$(k,m)=(-1,1)$ (filled circles) and $(-2,1)$ (open circles), computed in the traditional
approximation for the $5M_\odot$ ZAMS model for $\bar\Omega=0.4$. 
The $r$ modes associated with $(k,m)=(-1,1)$ and $(-2,1)$ correspond to
those with $l^\prime=m=1$ and $l^\prime=m+1=2$, respectively.}
\end{figure}

\begin{figure}
\centering
\epsfig{file=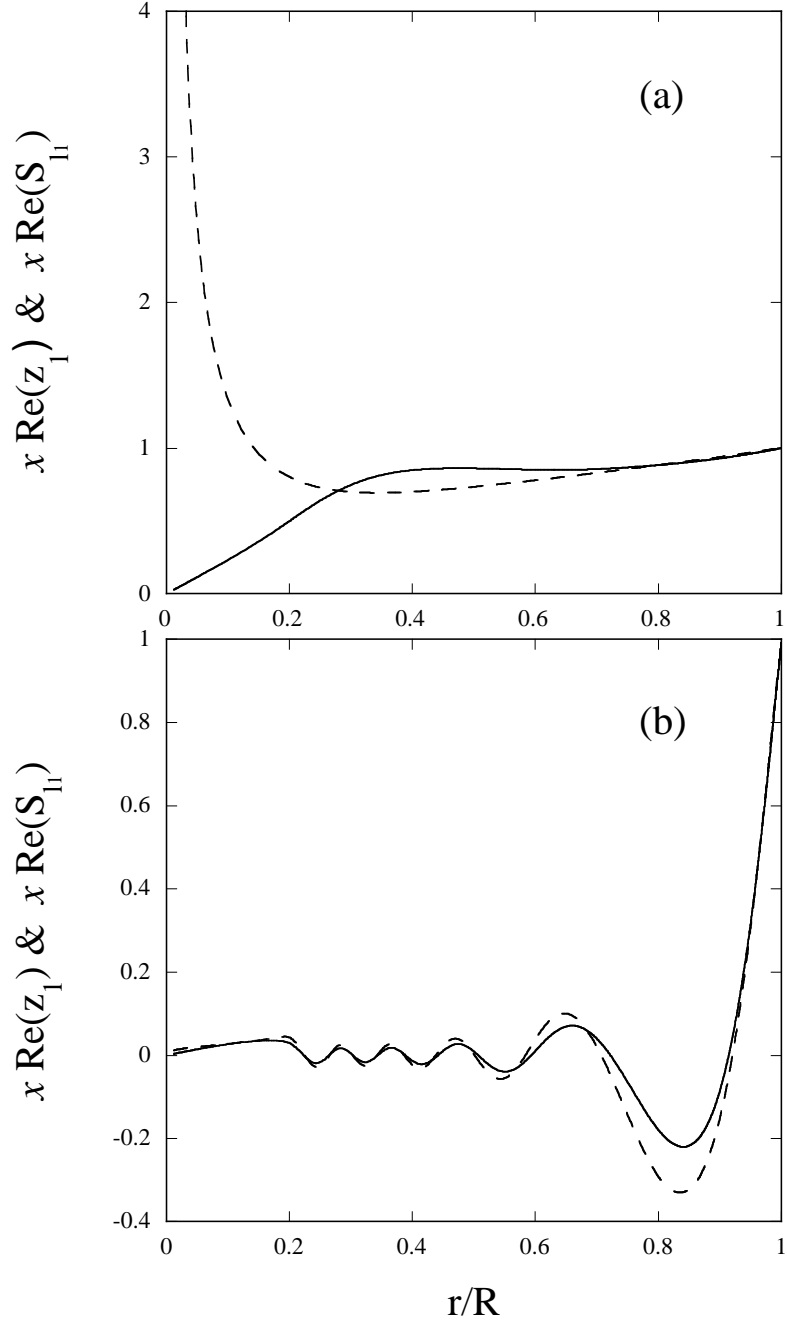,width=0.6\textwidth}
\caption{Eigenfunctions 
$x{\rm Re}(S_{l_1})$ (solid line) and $x{\rm Re}(z_1)$ (dashed line) 
are given versus $x=r/R$ for the 
fundamental $r_0$ modes with $l^\prime=m=1$ (panel a) and for 
$r_{10}$ mode with $l^\prime=m=1$ (panel b) 
of the $5M_\odot$ ZAMS model, where $\bar\Omega=0.4$.
The eigenfunctions $S_{l_1}$ and $z_1$ are computed without and with the
traditional approximation, respectively.
The amplitude normalization is given by ${\rm Re}(S_{l_1})=1$ and by ${\rm Re}(z_1)=1$
at the surface of the star. }
\end{figure}

\begin{figure}
\centering
\epsfig{file=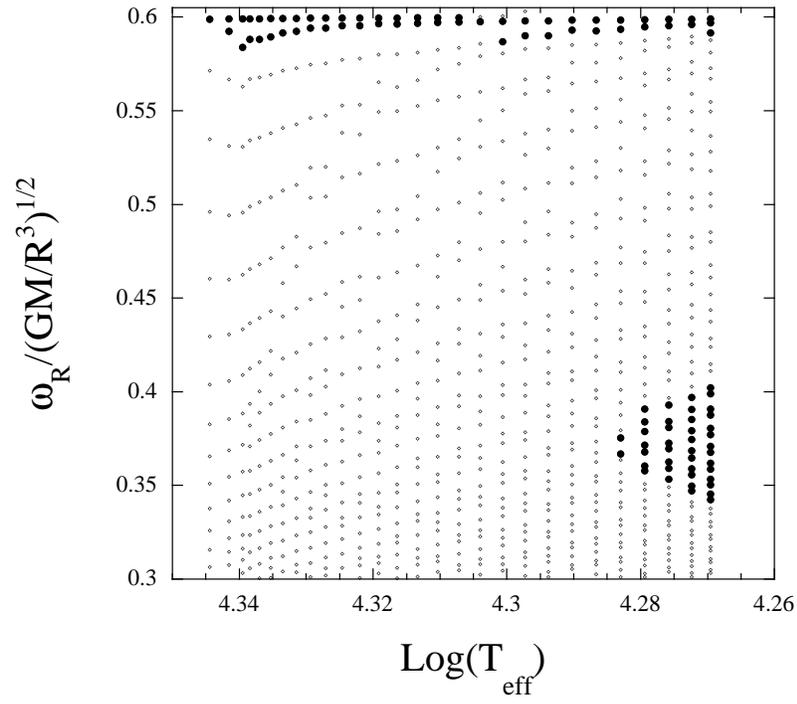,width=0.6\textwidth}
\caption{Same as Figure 9 but for $r$ modes computed in the traditional approximation.}
\end{figure}

\end{document}